\documentclass[aps,prl,twocolumn,superscriptaddress,showpacs,floatfix]{revtex4}
\usepackage{graphicx,epsfig,amsmath}

\newcommand{\bohrm}{\mbox{$\mu_{B}$}}
\newcommand{\kboltz}{\mbox{$k_{B}$}}

\newcommand{\etal}{\textit{et al.}}

\newcommand{\AlxGaAs}[2]{\mbox{$\text{Al}_{#1}\text{Ga}_{#2}\text{As}$}}
\newcommand{\Vds}{\mbox{$\text{V}_{\text{ds}}$}}
\newcommand{\Vg}{\mbox{$\text{V}_{\text{g}}$}}
\newcommand{\plaindidv}{\mbox{dI/d\Vds}}
\newcommand{\TKondo}{\mbox{$\text{T}_{\text{K}}$}}

\newcommand{\ZeemanE}{\mbox{$|g|\bohrm B$}}
\newcommand{\TMC}{\mbox{$\text{T}_{\text{MC}}$}}
\newcommand{\TFit}{\mbox{$\text{T}_{\text{Fit}}$}}
\newcommand{\DeltaK}{\mbox{$\Delta_{\text{K}}$}}
\newcommand{\subfig}[2]{Fig.~\ref{fig:#1}(#2)}

\begin{document}


\title{Measurements of Kondo and spin splitting in single-electron transistors}

\author{A. Kogan}
	\altaffiliation[Current address: ]{Department of Physics, University of Cincinnati, Cincinnati, OH 45221-0011}	
	\affiliation{Department of Physics, Massachusetts Institute of Technology, Cambridge, Massachusetts 02139}
\author{S. Amasha}
	\affiliation{Department of Physics, Massachusetts Institute of Technology, Cambridge, Massachusetts 02139}
\author{D. Goldhaber-Gordon}
	\altaffiliation[Current address: ]{Geballe Laboratory for Advanced Materials, McCullough Building, 
Room 346, 476 Lomita Mall, Stanford, California 94305-4045}
	\affiliation{Department of Physics, Massachusetts Institute of Technology, Cambridge, Massachusetts 02139}
\author{G. Granger}
	\affiliation{Department of Physics, Massachusetts Institute of Technology, Cambridge, Massachusetts 02139}
\author{M. A. Kastner} 
	\email{mkastner@mit.edu}
	\affiliation{Department of Physics, Massachusetts Institute of Technology, Cambridge, Massachusetts 02139}
\author{Hadas Shtrikman}
	\affiliation{Braun Center for Submicron Research, Weizmann Institute of Science, Rehovot, Israel 76100}


\begin{abstract}
	
We measure the spin splitting in a magnetic field $B$ of localized states in single-electron transistors using a new method, inelastic spin-flip cotunneling. Because it involves only internal excitations, this technique gives the most precise value of the Zeeman energy $\Delta = \ZeemanE$. In the same devices we also measure the splitting with $B$ of the Kondo peak in differential conductance. The Kondo splitting appears only above a threshold field as predicted by theory. However, the magnitude of the Kondo splitting at high fields exceeds $2 \ZeemanE$ in disagreement with theory.
   
\end{abstract}


\maketitle

	
	The Kondo state, formed when conduction electrons screen a magnetic impurity, has received new attention since its observation in nanostructures, because it is a spin-entangled state of the many-electron system. While the equilibrium properties of Kondo systems are described quantitatively by theory, non-equilibrium Kondo phenomena have proved harder to understand.
Single-electron transistors (SETs) provide the unique possibility of exploring these non-equilibrium phenomena. An SET consists of a confined droplet of electrons, called an artificial atom or a quantum dot, coupled by tunnel barriers to two conducting leads, called the source and the drain.  The electrochemical potential of the dot, as well as the coupling between the dot and the leads, can be tuned by changing voltages on electrodes. The Kondo effect in SETs \cite{goldhaber98:NaturePaper,cronenwett98:ScienceKondo,glazman88:ResonantKondo,ng88:CoulombRepulsion} 
occurs because a dot in a non-zero spin state coupled to its leads is analogous to a magnetic impurity coupled to the electrons in a host metal. Kondo correlations develop because the dot spin is screened by the spins of electrons in the leads. In an SET the Kondo state can be studied out of equilibrium by applying a DC voltage \Vds\space  between the source and the drain.

	Non-equilibrium Kondo physics is probed by the measurement of the splitting of the Kondo peak in differential conductance as a function of magnetic field $B$ \cite{meir93:nonequil,costi00:kondoh,moore00:ksplit,rosch03:KondoPertTheory}. Meir \etal~\cite{meir93:nonequil} predict that in a magnetic field the Kondo peaks occur at $e\Vds= \pm\Delta$. Here $\Delta = \ZeemanE$ is the Zeeman energy for spin splitting and $\bohrm= 58$  \mbox{$\mu$}eV/T is the Bohr magneton. Cronenwett \etal~\cite{cronenwett98:ScienceKondo} have measured the splitting of a Kondo peak and have found good agreement with this prediction, using $g= -0.44$ for bulk GaAs. However, more recent calculations by Costi \cite{costi00:kondoh} predict that the Kondo peak splitting should appear only above a critical magnetic field, and Moore and Wen \cite{moore00:ksplit} predict that the splitting should be smaller than $2\Delta$ at all fields. 

	Here we report measurements comparing the Kondo peak splitting to $\Delta$ in SETs. Using electron addition spectroscopy 
and a new, more precise method, inelastic spin-flip cotunneling, 
we determine $\Delta$ and hence the $g$-factor of our SETs. We find that the Kondo splitting appears 
only above a critical field as predicted by theory. However, the magnitude of the Kondo splitting at high fields 
is noticeably \textit{larger} than $2 \Delta$, contrary to theoretical predictions.  
	
\begin{figure} 
\setlength{\unitlength}{1cm}
\begin{center}
\begin{picture}(8,8.75)(0,0)

\put(0,0){\includegraphics[width=8.0cm, keepaspectratio=true]{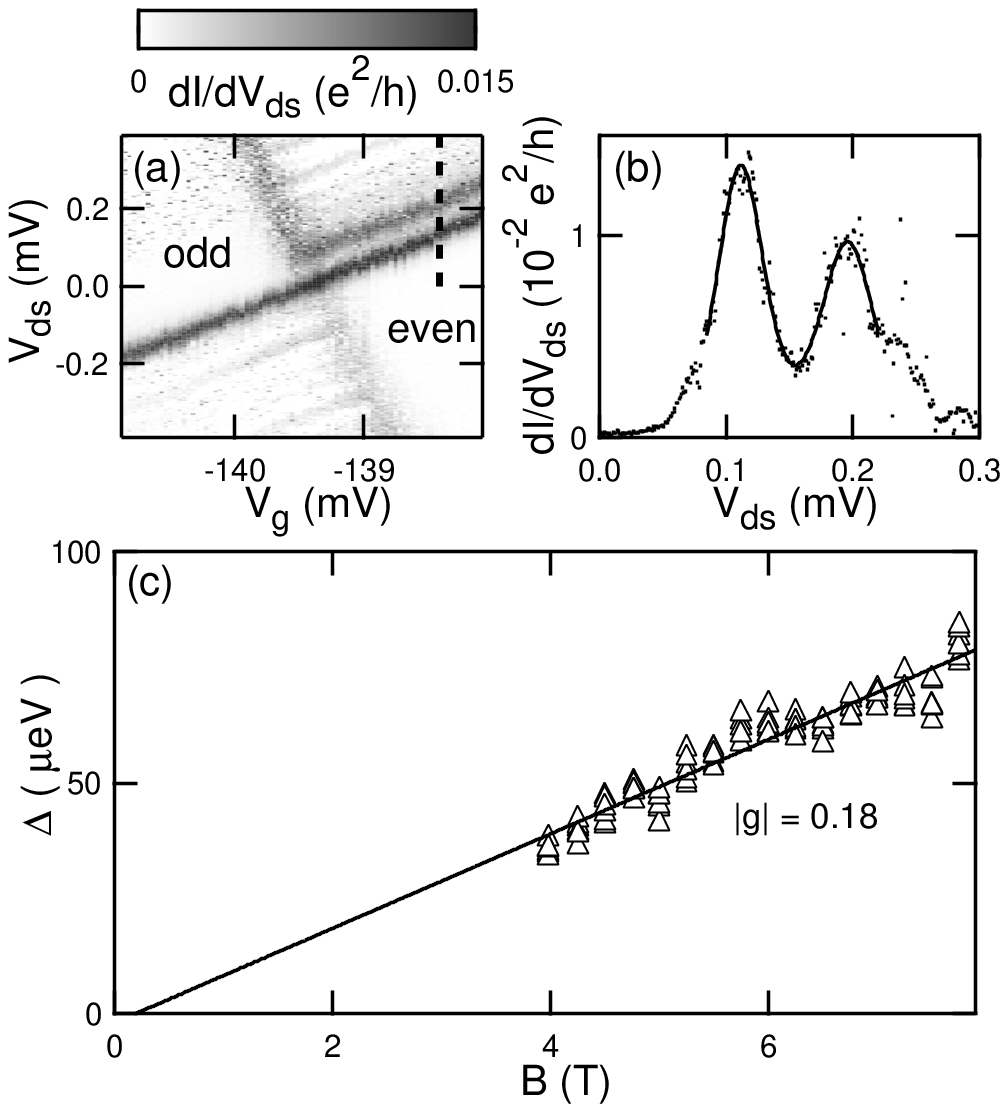}}
\put(1.4,2.2){\includegraphics[width=2.7cm, keepaspectratio=true]{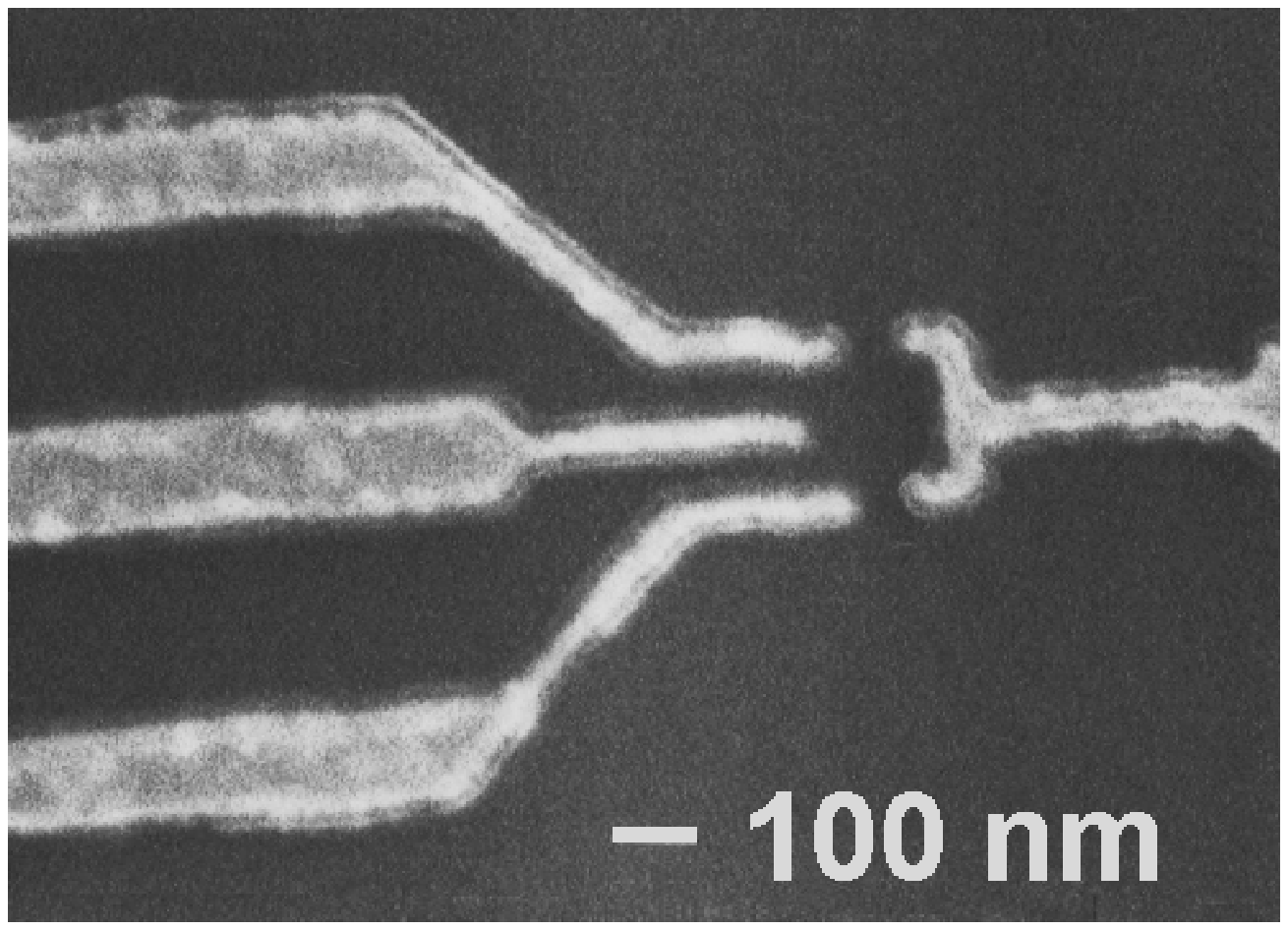}}

\put(1.1,3.125){\makebox(0,0){g}}
\put(1.25,3.125){\line(1,0){0.4}}

\end{picture}
\end{center}

\caption{(a) Differential conductance as a function of \Vds\space and \Vg\space at $B= 7.25$ T for SET 2. 
The odd and even Coulomb valleys are labeled. (b) Differential conductance as a function of 
drain-source voltage at the gate voltage marked by the dashed line in (a). The dots denote data while 
the solid line shows a fit to the sum of two $\cosh^{-2}$ functions. (c) Zeeman splitting as a function 
of magnetic field determined by fitting curves like that shown in (b). The solid line shows a fit to the 
data to determine $|g|$. The inset shows an electron micrograph of an SET similar to those we studied.}
\label{fig:excite}
\end{figure}

The two SETs we have studied are similar to those used by Goldhaber-Gordon 
\etal~\cite{goldhaber98:PRL,goldhaber98:NaturePaper}, and details about the AlGaAs/GaAs heterostructure can be found in the latter references. A two dimensional 
electron gas (2DEG) is formed at the AlGaAs/GaAs interface with an electron density of 
$8.1\times10^{11}$ $\text{cm}^{-2}$ and a mobility of $10^{5}$ $\text{cm}^{2}$/Vs at $4.2$ K. Magneto-transport shows that only one sub-band of the 2DEG is occupied.  Electron-beam lithography is used to define the gate electrodes shown in the inset of \subfig{excite}{c}. Applying a negative voltage to these electrodes depletes the 2DEG underneath them and forms an artificial atom of about $50$ electrons isolated by two tunnel barriers from the remaining 2DEG regions, the source and drain leads. The voltage on the gate electrode g is denoted \Vg.

	The two SETs were studied in different dilution refrigerators. SET 1 was measured in a Leiden 
Cryogenics $400$ $\mu$W dilution refrigerator with a $14$ T magnet and a lowest electron temperature of 
$20-30$ mK. The sample was aligned so that the 2DEG was parallel to the magnetic field to 
better than $0.5$ degree. SET 2 was measured in a $75$ $\mu$W Oxford Instruments dilution refrigerator 
with an $8$ T magnet and a lowest electron temperature of about $110$ mK. This sample could only 
be aligned parallel to the magnetic field to within a few degrees. To measure the differential conductance \plaindidv\space, we 
added a small sinusoidal modulation to \Vds\space and measured the resulting current with a preamplifier 
and a lock-in amplifier.
         
	One way to measure the Zeeman splitting of orbital levels in an SET is electron addition spectroscopy \cite{potok03:spincur,hanson03:spinrelaxPRL}. Figure \ref{fig:excite}(a) shows a plot of \plaindidv\space as a function of \Vds\space and \Vg\space while \subfig{excite}{b} shows \plaindidv\space at the value of \Vg\space marked on \subfig{excite}{a} with a dashed line, for which the occupancy of the dot is even. As we increase \Vds\space at this fixed \Vg, we lower the Fermi energy of the drain lead. The first peak in \plaindidv\space occurs when the Fermi energy of the drain aligns with the higher energy state of the spin split orbital state in the SET. Since even occupancy implies \textit{both} spin states are filled at $\Vds=0$, a second peak appears when the lower energy spin state enters the transport window. The separation of the two peaks is $\Delta / e(1-\alpha_{d})$, where $\alpha_{d}$ is the ratio of the dot-drain capacitance to the total capacitance. The $(1-\alpha_{d})$ factor arises because applying a positive voltage to the drain slightly lowers the energies of the levels in the dot. 
	
	To find the peak splitting we fit data like those in \subfig{excite}{b} to the sum of two 
$\cosh^{-2}$ functions; the specific form chosen does not affect the measurement of the splitting. Using 
the slopes of a Coulomb-blockade diamond, part of which is shown in \subfig{excite}{a}, we determine 
$\alpha_{d}= 0.15\pm 0.02$ and hence $\Delta$. Figure \ref{fig:excite}(c) shows $\Delta$ as a function of $B$. At each field we fit several traces like that in \subfig{excite}{b}. The scatter in the data arises because charge fluctuations near the SET are equivalent to gate voltage fluctuations and cause fluctuations in the peak positions. Fitting the data in \subfig{excite}{c} to a line gives $|g|= 0.18 \pm 0.04$. 

\begin{figure}[b]

\begin{center}
\includegraphics[width=7.0cm, keepaspectratio=true]{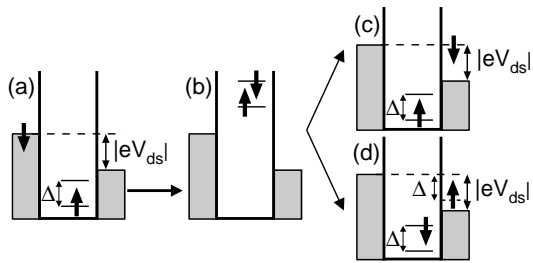}
\end{center}

	\caption{Cotunneling processes through a Zeeman split orbital, which can occur when the dot is in 
the Coulomb blockade regime with an odd number of electrons as in (a). We take spin up to be the lower 
energy state. A spin-down electron from the lead can tunnel onto the dot forming the virtual 
intermediate state shown in (b). For any value of \Vds, the spin-down electron can tunnel off the dot as 
shown in (c), resulting in elastic cotunneling. If $|e\Vds| \geq \Delta = \ZeemanE$, the spin-up 
electron can also tunnel off the dot resulting in inelastic cotunneling as shown in (d). 
}
	\label{fig:cotuncartoon}
\end{figure}

	We have developed a new, more precise method of measuring $\Delta$ using inelastic spin-flip cotunneling. 
Although resonant tunneling is prohibited in the Coulomb blockade regime, an electron can tunnel through 
the SET via a virtual intermediate state, a process known as cotunneling \cite{averin92:cotunneling}. 
Cotunneling can proceed elastically and leave the SET in its ground state, or inelastically, leaving it 
in an excited state. De Franceschi \etal~\cite{defranceschi01:orbitalcotun} observed inelastic 
cotunneling through an excited \textit{orbital} state of an SET with an even number of electrons. In our 
experiments the ground state has an odd number of electrons and the role of the excited state is played 
by the higher energy spin state. The elastic and inelastic cotunneling processes are illustrated in 
Fig.~\ref{fig:cotuncartoon}(a)-(d). If $|e\Vds| < \Delta$  only the elastic process is possible, but if 
$|e\Vds| \geq \Delta$ both the elastic and inelastic processes are possible and we expect steps in 
\plaindidv\space at $|e\Vds| = \Delta$. 

\begin{figure}

\begin{center}
\includegraphics[width=7.0cm, keepaspectratio=true]{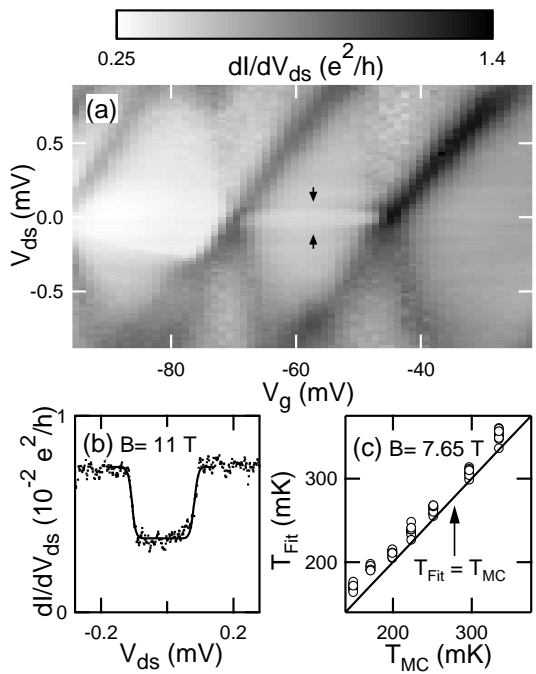}
\end{center}

\caption{
(a) Differential conductance as a function of \Vds\space and \Vg\space at $B= 8$ T for SET 1. The inelastic 
spin-flip cotunneling gap is in the middle of the center diamond and is marked by arrows. (b) Differential 
conductance as a function of drain-source voltage in the middle of the Coulomb valley for the same sample as 
in (a), but with different voltages on the electrodes and $B= 11$ T. The dots are the data and the solid 
line is a fit described in the text. (c) Temperatures extracted from fitting the cotunneling gap of SET 2 
at $B= 7.65$ T. The horizontal axis is the mixing chamber temperature \TMC\space and the vertical 
axis is the temperature \TFit\space extracted from fitting the gap as discussed in the text. The line shows 
$\TMC=\TFit$. 
}
	\label{fig:diamonds}
\end{figure}

\begin{figure}[!]
	\begin{center}
		\includegraphics[width=7.0cm, keepaspectratio=true]{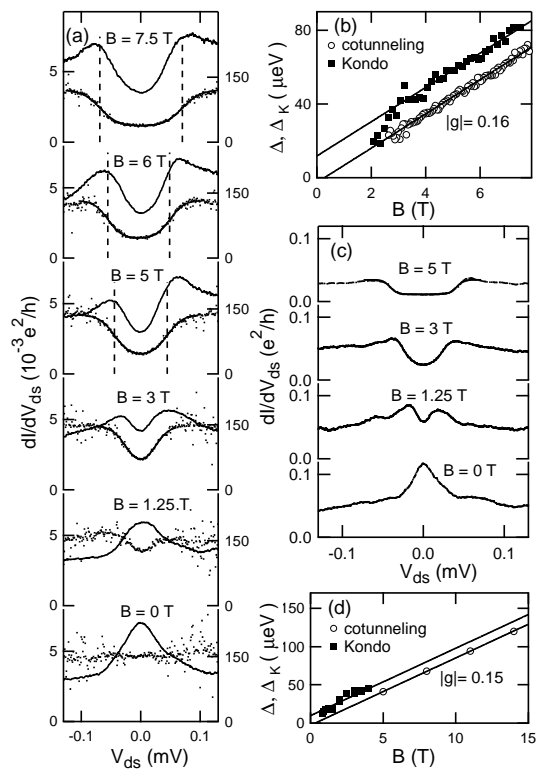}
	\end{center}
	\caption{(a) Evolution of a cotunneling gap (dots) and a Kondo peak (solid line) with increasing 
magnetic field in SET 2. The vertical scale on the left (right) is for the cotunneling data (Kondo 
data). For the cotunneling data, the coupling to the leads was reduced so that no Kondo peak was seen at 
zero field at the measurement temperature $T\approx 110$ mK. The solid lines through the cotunneling 
data at high fields are examples of fits to equation \ref{eq:lineshape}. The dashed lines at the three 
highest fields mark the center of the cotunneling edge; the lines are $2\Delta$ apart. (b) Splittings 
from cotunneling gaps (open circles) and Kondo peaks (solid squares) from SET 2. The solid line through 
the cotunneling data is a linear fit. The line through the Kondo data is the result of fixing the slope 
at the value obtained from the cotunneling data and fitting the data above $3$ T to extract the 
intercept. (c) Evolution of a Kondo peak into a cotunneling gap with increasing magnetic field in SET 1. 
The solid line in the $5$ T trace shows a fit as described in the text. (d) Splittings from the 
cotunneling gaps and Kondo peaks for SET 1. The Kondo splitting is determined at low field and the 
cotunneling gap at high field, but all other parameters are unchanged. We fit the data with lines the 
same way as for (b).
}	
	\label{fig:CotunNKondo}
\end{figure}

	This behavior is illustrated in \subfig{diamonds}{a} and (b). Note that the spin-flip cotunneling gap is independent of gate voltage and, as shown in \subfig{CotunNKondo}{a}, is only observed in a magnetic field. This is in contrast to cotunneling features associated with excited orbital states, like those in the Coulomb valley on the far left in Fig. 3(a), which depend on \Vg\space because changing the gate voltage alters the shape of the confining potential \cite{defranceschi01:orbitalcotun,kogan03:singlettriplet}. The orbital cotunneling features are also present at $B=0$.  To our knowledge, this is the first observation of the threshold for spin-flip cotunneling in an SET.

	Increasing the temperature broadens the inelastic cotunneling threshold. Lambe and Jaklevic 
\cite{Lambe68:molectun} have shown that for an inelastic process with negligible intrinsic width the 
lineshape is given by
	
\begin{equation}
	\frac{\mbox{dI}}{\mbox{d\Vds}}= A_{e} + A_{i}[F\left(\frac{e\Vds+\Delta}{\kboltz \text{T}}\right)+ 
F\left(-\frac{e\Vds-\Delta}{\kboltz \text{T}}\right) ]
\label{eq:lineshape}
\end{equation}
In this equation, $A_{e}$ is the conductance from elastic cotunneling and $A_{i}$ describes the additional contribution 
from inelastic cotunneling. $F$ is a function defined by $	F(x)= [1+(x-1)\exp(x)]/[\exp(x)-1]^{2}$.	This lineshape has steps centered at $\Vds= \pm \Delta /e$ with width $5.4\kboltz\text{T}/e$. A fit to Eq. \ref{eq:lineshape} is shown as the solid line in \subfig{diamonds}{b} and is in excellent agreement with the data.  Figure \ref{fig:diamonds}(c) shows the temperature \TFit\space extracted from fits to data taken at different temperatures with SET 2. The sharpest cotunneling step that we have observed (shown in \subfig{diamonds}{b}) is for SET 1 and has width $22$ $\mu$eV. This can be used to place a lower bound on the spin decoherence time because the intrinsic width of the cotunneling step is inversely proportional to the decoherence time.   

	Figure \ref{fig:CotunNKondo}(a) shows the evolution with field of a cotunneling feature in SET 2. 
The splittings $\Delta$ from fits to Eq. \ref{eq:lineshape} are plotted as functions of $B$ 
in \subfig{CotunNKondo}{b}. Fitting these data to a line gives $|g|= 0.16 \pm 0.02$, consistent with the 
value from addition spectroscopy but with greater precision. Measurements for SET 1 in high 
magnetic fields, shown in Figs. \ref{fig:CotunNKondo}(c) and (d), give $|g|= 0.152 \pm 0.006$, consistent with the value for SET 2. The inelastic cotunneling method, which measures the spectrum of internal excitations of the dot, is more precise than electron addition spectroscopy for two reasons. First, the cotunneling spectrum is independent of the chemical potential of the dot, making the splitting insensitive to small charge fluctuations near the SET. Second, the cotunneling method is inherently more precise because its intrinsic width is smaller than the width of the charging peak, which is determined by the coupling to the leads and is approximately $\Gamma= 35\space\mu$eV for SET 2 in Fig.~\ref{fig:excite}. 

	The splitting of the Kondo peak at high fields is larger than the cotunneling gap, as can be seen, 
for example, from the data in \subfig{CotunNKondo}{a}. To compare our findings to theory, we follow Moore and Wen \cite{moore00:ksplit} who identify the field-induced splitting between the two peaks in their calculated spectral function with the splitting in \Vds\space between the peaks in \plaindidv. We define $\DeltaK/e$ as half the separation in \Vds\space between the two peaks. Figure \ref{fig:CotunNKondo}(b) shows \DeltaK\space for SET 2 from data like those shown in \subfig{CotunNKondo}{a}. There is a clear offset between \DeltaK\space and the cotunneling data at high fields; fitting the data above $3$ T to a line with $|g|$ fixed at the value obtained from cotunneling gives an offset of $12 \pm 1$ $\mu$eV. 

	The Kondo peak evolves smoothly into a cotunneling threshold with increasing $B$ when there is an odd number of electrons in the dot. Figure \ref{fig:CotunNKondo}(c) shows this evolution for SET 1. Note that the values of $\Delta$ in \subfig{CotunNKondo}{d} are extracted from the high-field data for which the Kondo peak is absent. The \DeltaK\space are extracted from the low-field data and are plotted in \subfig{CotunNKondo}{d}. We fit the data to a line with the slope fixed at the value obtained from the cotunneling data points, $|g|= 0.152$. The fit gives a y-intercept of $10 \pm 2$ $\mu$eV, similar to the result for SET 2. Note that in contrast to the Kondo data, both the addition spectroscopy and cotunneling measurements extrapolate to zero within the errors at $B= 0$. Contrary to theoretical predictions \cite{moore00:ksplit}, \DeltaK\space is \textit{larger} than \ZeemanE. 	

	Another feature of the Kondo data is that the splitting only appears above a threshold magnetic field. For example, the $1.25$ T data in \subfig{CotunNKondo}{a} shows no evidence of splitting. The data for \DeltaK\space in \subfig{CotunNKondo}{b} shows a threshold for splitting near $2$ T, which is where
$\ZeemanE/\kboltz \approx \TKondo \approx 300$ mK.  Costi \cite{costi00:kondoh} predicts a threshold field for 
the Kondo splitting. In this respect our results are consistent with theory.
	
	The value of $|g|$ in our SETs is $0.16$, much smaller 
than the bulk GaAs value of $0.44$. The small g-factor may arise because the electron density in our SETs is much larger than in typical devices. Variations in the g-factor with density have previously been observed in AlGaAs/GaAs heterostructures \cite{stein83:ESRonAlGaAs_GaAs,dobers88:ESRin2DEG,snelling91:ginGaAs_AlGaAs,jiang01:gateg,Salis01:gfactorControl}. We believe our small value results from penetration of the electrons into the \AlxGaAs{0.3}{0.7} which has $g= +0.4$, and the large 2DEG density in our SETs enhances this effect. Linearly extrapolating the results of Jiang and Yablonovitch \cite{jiang01:gateg}  gives a change in $|g|$ that is about half of that we observe. We plan a more realistic calculation to test this hypothesis more carefully.

	The observation that $\DeltaK > \ZeemanE$ is surprising.  Moore and Wen \cite{moore00:ksplit} predict that $\DeltaK<\ZeemanE$ at all $B$, whereas we find that $\DeltaK > \ZeemanE$ by 
$\approx 10$ $\mu$eV at high fields. Our results differ from those of Goldhaber-Gordon \etal~\cite{goldhaber98:NaturePaper}. However, the measurements reported by Goldhaber-Gordon \etal~are taken with the magnetic field normal to the 2DEG and could involve effects from the orbital motion of the electrons. Such effects are not considered by Moore and Wen \cite{moore00:ksplit} and are eliminated in the present experiments by orienting the 2DEG parallel to the magnetic field. One might ask if elastic cotunneling forms a background that could explain our observations. Using zero field data like those in \subfig{CotunNKondo}{a}, only over a wider range of \Vds, we fit the cotunneling conductance outside the Kondo peak region to a parabola and subtract this from our Kondo splitting data from SET 2. This decreases the high-field offset by about $5$\%, which does not affect our conclusions. However, such a subtraction is not well motivated since elastic and inelastic cotunneling processes contribute coherently to the Kondo correlations. Thus the observation that $\DeltaK > \ZeemanE$ is not understood and suggests that more theoretical analysis of the non-equilibrium Kondo problem is important.

	We are grateful to J. Moore, L. Levitov, X.-G. Wen, W. Hofstetter, and J. Folk for discussions and to C. Cross for experimental help. We thank D. Mahalu for electron-beam lithography. We acknowledge support from the NSERC of Canada (G. G.). This work was supported by the US Army Research Office under Contract DAAD19-01-1-0637, by the National Science Foundation under Grant No.~DMR-0102153, and in part by the NSEC Program of the National Science Foundation under Award Number DMR-0117795 and the MRSEC Program of the National Science Foundation under award number DMR 02-13282.


\begin{thebibliography}{16}

\expandafter\ifx\csname natexlab\endcsname\relax\def\natexlab#1{#1}\fi
\expandafter\ifx\csname bibnamefont\endcsname\relax
  \def\bibnamefont#1{#1}\fi
\expandafter\ifx\csname bibfnamefont\endcsname\relax
  \def\bibfnamefont#1{#1}\fi
\expandafter\ifx\csname citenamefont\endcsname\relax
  \def\citenamefont#1{#1}\fi
\expandafter\ifx\csname url\endcsname\relax
  \def\url#1{\texttt{#1}}\fi
\expandafter\ifx\csname urlprefix\endcsname\relax\def\urlprefix{URL }\fi
\providecommand{\bibinfo}[2]{#2}
\providecommand{\eprint}[2][]{\url{#2}}

\bibitem{goldhaber98:NaturePaper}
\bibinfo{author}{\bibfnamefont{D.}~\bibnamefont{Goldhaber-Gordon}} \textit{et~al.},
  \bibinfo{journal}{Nature} \textbf{\bibinfo{volume}{391}},
  \bibinfo{pages}{156} (\bibinfo{year}{1998}{\natexlab{a}}).

\bibitem{cronenwett98:ScienceKondo}
\bibinfo{author}{\bibfnamefont{S.}~\bibnamefont{Cronenwett}} \textit{et~al.},
  \bibinfo{journal}{Science} \textbf{\bibinfo{volume}{281}},
  \bibinfo{pages}{540} (\bibinfo{year}{1998}).

\bibitem[{\citenamefont{Glazman and Raikh}(1988)}]{glazman88:ResonantKondo}
\bibinfo{author}{\bibfnamefont{L.~I.} \bibnamefont{Glazman}} \bibnamefont{and}
  \bibinfo{author}{\bibfnamefont{M.~E.} \bibnamefont{Raikh}},
  \bibinfo{journal}{JETP Lett.} \textbf{\bibinfo{volume}{47}},
  \bibinfo{pages}{452} (\bibinfo{year}{1988}).

\bibitem[{\citenamefont{Ng and Lee}(1988)}]{ng88:CoulombRepulsion}
\bibinfo{author}{\bibfnamefont{T.~K.} \bibnamefont{Ng}} \bibnamefont{and}
  \bibinfo{author}{\bibfnamefont{P.~A.} \bibnamefont{Lee}},
  \bibinfo{journal}{Phys. Rev. Lett.} \textbf{\bibinfo{volume}{61}},
  \bibinfo{pages}{1768} (\bibinfo{year}{1988}).

\bibitem{meir93:nonequil}
\bibinfo{author}{\bibfnamefont{Y.}~\bibnamefont{Meir}} \textit{et~al.},
  \bibinfo{journal}{Phys. Rev. Lett.} \textbf{\bibinfo{volume}{70}},
  \bibinfo{pages}{2601} (\bibinfo{year}{1993}).

\bibitem[{\citenamefont{Costi}(2000)}]{costi00:kondoh}
\bibinfo{author}{\bibfnamefont{T.~A.}~\bibnamefont{Costi}},
  \bibinfo{journal}{Phys. Rev. Lett.} \textbf{\bibinfo{volume}{85}},
  \bibinfo{pages}{1504} (\bibinfo{year}{2000}).

\bibitem[{\citenamefont{Moore and Wen}(2000)}]{moore00:ksplit}
\bibinfo{author}{\bibfnamefont{J.~E.} \bibnamefont{Moore}} \bibnamefont{and}
  \bibinfo{author}{\bibfnamefont{X.-G.} \bibnamefont{Wen}},
  \bibinfo{journal}{Phys. Rev. Lett.} \textbf{\bibinfo{volume}{85}},
  \bibinfo{pages}{1722} (\bibinfo{year}{2000}).

\bibitem{rosch03:KondoPertTheory}
\bibinfo{author}{\bibfnamefont{A.}~\bibnamefont{Rosch}} \textit{et~al.},
  \bibinfo{journal}{Phys. Rev. Lett.} \textbf{\bibinfo{volume}{90}},
  \bibinfo{pages}{076804} (\bibinfo{year}{2003}).

\bibitem{goldhaber98:PRL}
\bibinfo{author}{\bibfnamefont{D.}~\bibnamefont{Goldhaber-Gordon}} \textit{et~al.},
  \bibinfo{journal}{Phys. Rev. Lett.} \textbf{\bibinfo{volume}{81}},
  \bibinfo{pages}{5225} (\bibinfo{year}{1998}{\natexlab{b}}).

\bibitem{potok03:spincur}
\bibinfo{author}{\bibfnamefont{R.}~\bibnamefont{Potok}} \textit{et~al.},
  \bibinfo{journal}{Phys. Rev. Lett.} \textbf{\bibinfo{volume}{91}},
  \bibinfo{pages}{016802} (\bibinfo{year}{2003}).

\bibitem{hanson03:spinrelaxPRL}
\bibinfo{author}{\bibfnamefont{R.}~\bibnamefont{Hanson}} \textit{et~al.},
  \bibinfo{journal}{Phys. Rev. Lett.} \textbf{\bibinfo{volume}{91}},
  \bibinfo{pages}{196802} (\bibinfo{year}{2003}).

\bibitem[{\citenamefont{Averin and Nazarov}(1992)}]{averin92:cotunneling}
\bibinfo{author}{\bibfnamefont{D.~V.} \bibnamefont{Averin}} \bibnamefont{and}
  \bibinfo{author}{\bibfnamefont{Yu.~V.} \bibnamefont{Nazarov}}, in
  \emph{\bibinfo{booktitle}{Single Charge Tunneling}}, edited by
  \bibinfo{editor}{\bibfnamefont{H.}~\bibnamefont{Grabert}} \bibnamefont{and}
  \bibinfo{editor}{\bibfnamefont{M.~H.}~\bibnamefont{Devoret}}
  (\bibinfo{publisher}{Plenum Press}, \bibinfo{year}{1992}), NATO ASI Series B
  294, pp. \bibinfo{pages}{217--246}.

\bibitem{defranceschi01:orbitalcotun}
\bibinfo{author}{\bibfnamefont{S.}~\bibnamefont{{De Franceschi}}} \textit{et~al.},
  \bibinfo{journal}{Phys. Rev. Lett.} \textbf{\bibinfo{volume}{86}},
  \bibinfo{pages}{878} (\bibinfo{year}{2001}).

\bibitem{kogan03:singlettriplet}
\bibinfo{author}{\bibfnamefont{A.}~\bibnamefont{Kogan}} \textit{et~al.},
  \bibinfo{journal}{Phys. Rev. B} \textbf{\bibinfo{volume}{67}},
  \bibinfo{pages}{113309} (\bibinfo{year}{2003}).

\bibitem[{\citenamefont{Lambe and Jaklevic}(1968)}]{Lambe68:molectun}
\bibinfo{author}{\bibfnamefont{J.}~\bibnamefont{Lambe}} \bibnamefont{and}
  \bibinfo{author}{\bibfnamefont{R.~C.} \bibnamefont{Jaklevic}},
  \bibinfo{journal}{Phys. Rev.} \textbf{\bibinfo{volume}{165}},
  \bibinfo{pages}{821} (\bibinfo{year}{1968}).

\bibitem{stein83:ESRonAlGaAs_GaAs}
\bibinfo{author}{\bibfnamefont{D.}~\bibnamefont{Stein}} \textit{et~al.},
  \bibinfo{journal}{Phys. Rev. Lett.} \textbf{\bibinfo{volume}{51}},
  \bibinfo{pages}{130} (\bibinfo{year}{1983}).

\bibitem{dobers88:ESRin2DEG}
\bibinfo{author}{\bibfnamefont{M.}~\bibnamefont{Dobers}} \textit{et~al.},
  \bibinfo{journal}{Phys. Rev. B} \textbf{\bibinfo{volume}{38}},
  \bibinfo{pages}{5453} (\bibinfo{year}{1988}).

\bibitem{snelling91:ginGaAs_AlGaAs}
\bibinfo{author}{\bibfnamefont{M.J.}~\bibnamefont{Snelling}} \textit{et~al.},
  \bibinfo{journal}{Phys. Rev. B} \textbf{\bibinfo{volume}{44}},
  \bibinfo{pages}{11345} (\bibinfo{year}{1991}).

\bibitem[{\citenamefont{Jiang and Yablonovitch}(2001)}]{jiang01:gateg}
\bibinfo{author}{\bibfnamefont{H.~W.} \bibnamefont{Jiang}} \bibnamefont{and}
  \bibinfo{author}{\bibfnamefont{E.}~\bibnamefont{Yablonovitch}},
  \bibinfo{journal}{Phys. Rev. B} \textbf{\bibinfo{volume}{64}},
  \bibinfo{pages}{041307(R)} (\bibinfo{year}{2001}).

\bibitem{Salis01:gfactorControl}
\bibinfo{author}{\bibfnamefont{G.}~\bibnamefont{Salis}} \textit{et~al.},
  \bibinfo{journal}{Nature} \textbf{\bibinfo{volume}{414}},
  \bibinfo{pages}{619} (\bibinfo{year}{2001}).


\end{thebibliography}

\end{document}